\documentclass[preprint,12pt]{elsarticle}
\usepackage{float}
\usepackage{psfrag}
\usepackage{amsmath}
\usepackage{amsfonts}
\usepackage{graphicx}
\usepackage{color}
\usepackage[utf8]{inputenc}
\usepackage{lscape}
\usepackage{multicol}
\usepackage{srcltx}
\usepackage{overpic}
\usepackage{titletoc}
\usepackage{multirow}
\usepackage{tikz}
\usepackage{enumitem}
\usepackage[papersize={216mm,279mm},tmargin=20mm,bmargin=30mm,lmargin=30mm,rmargin=30mm]{geometry}

\journal{J. Phys.: Cond. Matter}

\begin{document}
\markboth{J. Calle, D. Sanjin\'es and E. Mamani}{mass tensor in the lattice}

\title{Dynamical behaviour of the mass tensor in the lattice}
\author{J. Calle, D. Sanjin\'es, E. Mamani}
\address{Carrera de F\'isica, Universidad Mayor de San Andr\'es, c. 27 Cota - Cota, casilla de correos 8635, La Paz, Bolivia.\\email: diegosanjinescastedo@gmail.com}

\begin{abstract}
We study with the semiclassical model the dynamics of a particle moving in a orthorrombic lattice by the hopping mechanism in the presence of a static Coulomb field and a rapidly oscillating homogeneous field. Based on results for the anisotropic Kepler problem, where a chaotic trajectory arises due to different masses of the particle in the X and Y directions, and since this result is also obtained in the rectangular lattice at the continuum limit, the application of the oscillating field allows reparameterizing the effective masses so that they become equal in both directions. Thus, the result is an elliptic trajectory in the continuum limit of the lattice. This chaos--order transition is confirmed by the calculation of Lyapunov exponents for the different trajectories. It is also found that the application of an external magnetic field in the Z direction of the lattice yields chaos control which can be generalized beyond the simple case of the circular trajectory shown in this work. The previous scheme of chaos control was extended for trajectories in 3D in the continuum limit of the orthorrombic lattice which is also equivalent to the far region approximation where the particle moves with respect to the Coulomb field origin. In this case it found that the chaos control technique permits to observe two different dynamical regimes which, although consistent with Bohr-van Leeuwen theorem, suggests that an observable way to distinguish one from the other is to be sought.
\end{abstract}


\maketitle

\section{Introduction}

The anisotropic Kepler problem goes back to the description of a system of central forces of two
particles on a semiconductor wafer. One particle is a positive charge ion at rest (donor impurity) and the other is a negative charge particle (electron), which moves around the impurity \cite{Luttinger1954}. This inspired studies of the classical anisotropic Kepler problem, where a particle in a central force field with the classical Hamiltonian is studied (notably by Gutzwiller M. C. and referred to by Reichl L. \cite{Reichl2021} ), obtaining a chaotic trajectory by defining the existence of a different mass parameter for each axis. Thus, and according to a recent study of the Kepler problem \cite{Sanjines2025} where we investigated the semiclassical dynamics of a spinless charged particle subjected to a Coulomb potential $U(x,y,z)=-U_1/(x^2+y^2+z^2)^{1/2}$, the mass tensor anisotropy arises naturally, leading to chaotic trajectories in the continuum limit of the rectangular lattice wherein the lattice constants tend to zero but their ratio is maintained. Moreover, it is shown that the mass tensor anisotropy disappears if the Kepler problem is studied on a square lattice, yielding stationary trajectories (quasi-ellipses) and ellipses in the continuum limit, since angular momentum is conserved due to the rotational symmetry (isotropy) of space.

There are results that consider rapidly oscillating external electric fields in the
semiclassical tight-binding model \cite{Mamani2017}; thus, by performing the temporal averaging method of Kapitza referred to by Landau L. D. \cite{Landau}, an effective time-independent Hamiltonian is deduced with hopping elements that depend on the field parameters. By manipulating such parameters (components of the field), it was achieved, for example, the suppression of the nearest-neighbor hopping elements, leaving the second-neighbor interaction dominant. This is what we understand in \cite{Mamani2017} as the ``duplication" of the lattice applying interaction engineering. The generalization of the previous results was achieved in a rectangular lattice, also rendering the 2D lattice duplication effect.\footnote{Delgado, S. C., \textit{Duplicaci\'on de la red rectangular con ingenier\'ia de interacciones} (proyecto de grado 2025), Carrera de F\'isica, Universidad Mayor de San Andr\'es, La Paz, Bolivia.} With the results of the aforementioned works, it is known that the application and manipulation of a rapidly oscillating electric field can modify the dynamics of the particle moving in a lattice through the reparameterized components of the effective mass tensor.

In this work the Kepler problem was addressed in a orthorrombic lattice with the tight-binding model to describe the dynamics of the charged spinless particle \textit{via} the hopping mechanism in the semiclassical approximation. It should be emphasized that the strict way to treat this problem is through quantum
mechanics. However, in \cite{Velasco2021} the dynamics of a charged particle in a one-dimensional lattice under the effect of a Coulomb potential due to an impurity is studied, where it is shown that quantum results coincide reasonably with those obtained by the semiclassical model in the continuum. Such manipulation was extended within the continuum limit of the rectangular lattice, thus achieving a diagonal mass tensor with different elements, yielding in consequence a chaotic particle's trajectory. The application of a rapidly oscillating external field allows these elements of the mass tensor to be equalized, so that the trajectory of the particle becomes an ellipse, as it should be the solution for the trajectory in the Kepler problem in an isotropic space.

Our work in the next sections is organized as follows: In Section 2 we develop the semiclassical model (Hamilton equations) for the orthorrombic lattice. In Section 3 we subject the elements of the mass tensor to an external electric field so as to render those elements equal (by the \textit{interactions engineering} technique). In Section 4 we carry out the specific application of chaos control in the continuum limit of the lattice; we test the results by the variation of the Lyapunov exponent. In Section 5 we apply the interactions engineering scheme to achieve chaos control in the rectangular lattice (without invoquing the continuum limit) for the na\"{\i}ve case of the circular orbit in a homogeneous external magnetic field. In Section 6 we extended to the 3D lattice the criteria for chaos control; as a consequence we can observe two dynamical regimes which -in principle- should be distinguished one from the other by any observable property at the impurity site. In Section 7 we draw some overall relevant conclusions of our work and point a few perpectives.

\section{Semiclassical model for the orthorrombic lattice}

The tight-binding Hamiltonian function for an electron moving by the hopping mechanism in a 3-dimensional orthorhombic lattice with lattice constants $a, b, c$ along the $X, Y, Z$ axes, with hopping elements $A, B, C$ respectively, and with a potential energy due to an external static force, is

\begin{align}
H(\textbf{r},\textbf{k})=&2A(1-\cos{ak_x})+2B(1-\cos{bk_y})\nonumber\\&+2C(1-\cos{ck_z})+U(\textbf{r}).\label{1}
\end{align}This function was constructed according to the semi-classical method (as shown, for example, in \cite{Ashcroft}) from the tight-binding band for the hopping particle moving in a periodic potential with a position vector in the Bravais lattice $\textbf{R}=(na,mb,qc)$ for
integers $n,m,q$ and a quasi-momentum vector $\hbar\textbf{k}=\hbar (k_x,k_y,k_z)$. In the limit of the effective mass approximation, where $m_x=\lim_{A\rightarrow \infty, a\rightarrow 0} \hbar^2/(2Aa^2)$, the terms $\hbar k_x$ and $\hbar^2 k_x^2/(2 m_x)$ are the momentum and energy, respectively, of the free-particle along the $X$ axis. And similarly for the $Y$ and $Z$ axes.

Thus, the classical equations of motion of Hamilton (with $\textbf{p}=\hbar \textbf{k}$) are, for the time variation of the coordinates,

\begin{align}
\hbar \dot{x}&=\frac{\partial H}{\partial k_{x}}=(2Aa/\hbar)\sin{ak_{x}},\label{2}\\
\hbar \dot{y}&=\frac{\partial H}{\partial k_{y}}=(2Bb/\hbar)\sin{bk_{y}},\label{3}\\
\hbar \dot{z}&=\frac{\partial H}{\partial k_{z}}=(2Cc/\hbar)\sin{ck_{z}};\label{4}
\end{align}and, for the time variation of momenta,
\begin{align}
\hbar \dot{k}_{x}&=-\frac{\partial H}{\partial x}=-\frac{\partial U}{\partial x},\label{5}\\
\hbar \dot{k}_{y}&=-\frac{\partial H}{\partial y}=-\frac{\partial U}{\partial y},\label{6}\\
\hbar \dot{k}_{z}&=-\frac{\partial H}{\partial z}=-\frac{\partial U}{\partial z}.\label{7}
\end{align}These equations, together with their corresponding initial conditions at $t=0$ for the position $\textbf{r}(0)$ and for the velocity $\textbf{v}(0)=\dot{\textbf{r}}(0)$, constitute the semiclassical model of the tight-binding dynamics for the hopping particle in the lattice. Choosing the initial conditions such that the particle moves only in the $XY$ plane: $z(0)=\dot{z}(0)=0$, and for an energy potential such that $\hbar \dot{k}_z(0)=-\partial U/\partial z|_{z=0}=0$ in (\ref{7}), it follows from (\ref{4}) that $k_z(0)=0$; then it holds that $\ddot{z}(0)=2Cc^2\dot{k}_z(0)\cos(ck(0))=0$, i.e., the particle will continue moving in the $XY$ plane in the absence of any perpendicular acceleration that could drag it out of the plane.\footnote{This would seem a rather obvious conclusion, but the same argument applies to a particle whose initial conditions do not lie on the XY plane. In the continuum limit the trajectories are ellipses similar to those of Fig.\;\ref{fig9} if the chaos control conditions for the external electric field are met (eq. (\ref{16}) below); otherwise, chaotic trajectories occur which lie on the oblique plane defined by the initial position and momentum vectors.} Therefore we will consider hereafter the dynamics of an electron in a rectangular lattice described by equations (\ref{2}), (\ref{3}), (\ref{5}) and (\ref{6}) above (with $z=0$). For simplicity of notation, a unit value will be taken for the angular Planck constant, $\hbar=1$. This implies that momentum and energy should be expressed in units of 1/meter and 1/second, respectively,but the choice of $\hbar=1$ in this work does not entail any specific physical meaning, as its true magnitude can be introduced in the results before interpreting the physical measurements.

Next we invoke the mechanism of \textit{interactions engineering} applied to a tight-binding rectangular
lattice described, for example, in \cite{Mamani2017}: if the charged particle moving in the lattice by the hopping mechanism is subjected to an external force caused by a homogeneous and rapidly oscillating electric field $\textbf{f} (\omega t)=2(fx,fy)\cos(\omega t)$, then the Hamiltonian function (\ref{1}) is augmented by a
corresponding potential energy,

\begin{align}
H(\textbf{r},\textbf{k};t)=H(\textbf{r},\textbf{k})+\textbf{r} \cdot \textbf{f} (\omega t),\label{1a}
\end{align}which admits the canonical transformation $H(\textbf{r},\textbf{k};t) \rightarrow H(\textbf{r},\textbf{k}'(t))$, where $\textbf{k}'(t)=\textbf{k}+\int^t \textbf{f} (\omega t') dt'$. From this latter form of the Hamiltonian an effective time-independent Hamiltonian is obtained by Kapitza's time-averaging method \cite{Landau}:

\begin{equation}
H_{\text{ef}}=-2\tilde{A}\cos{aK_x}-2\tilde{B}\cos{2bK_y}+U(X,Y),\label{8}
\end{equation}where

\begin{equation}
\tilde{A}= A J_0(2af_x/\omega),
\tilde{B}= B J_0(2bf_y/\omega),\label{9}
\end{equation}are the effective hopping elements that depend on the parameters of the rapidly oscillating external field, and $J_0$ is the Bessel function of the first kind of order zero. Thus, equations (\ref{2})-(\ref{7}) hold for the effective dynamic variables $X,Y,Z,K_x,K_y,K_z$ instead of $x,y,z,k_x,k_y,k_z$, and with $H_{\text{ef}}$ instead of $H(\textbf{r},\textbf{k})$. By adding a homogeneous external field of according to (8), the system exchanges energy with the field with a period of $T= 2\pi/\omega$, and therefore the energy is not constant. When $\omega$ is very large, the Kapitza time-averaging method \cite{Landau} is applied to obtain the effective Hamiltonian $H_{ef}$ in \eqref{8}, which no longer depends on time, so the system's energy is constant.

\section{Effective mass tensor}

Under the scheme of interactions engineering described in the last section, the effective mass tensor $m(\textbf{K})$ in the rectangular lattice is reparameterized due to the action of the homogeneous external field $\textbf{f}(\omega t)$. Its inverse components are given by \cite{Ashcroft}:

\begin{equation}
\left[\left(m\right)^{-1}\right]_{ij} = \frac{1}{\hbar^2}
\frac{\partial^2 H}{\partial K_i \, \partial K_j}.\label{10}
\end{equation}In the following sections this new result will be applied to the reparameterized effective mass in a rectangular lattice with a Coulomb potential $U(r)=-U_1/r$ in which the chaotic trajectories of the electron disappear and give rise to ellipses (Kepler problem), which suggests a kind of chaos control by a rapidly oscillating external field.

When considering the dynamics of an electron in a rectangular lattice, we can explain naturally the anisotropy of
the mass tensor by calculating its components according to \eqref{10} applying the tight-binding Hamiltonian \eqref{1}:

\begin{align}
\label{11}
m_x&=\left[2Aa^2\cos(aK_x(t))\right]^{-1},\nonumber\\
m_y&=\left[2Bb^2\cos(bK_y(t))\right]^{-1}.
\end{align}The ratio of these effective masses in the continuum is

\begin{align}
\label{12}
\frac{m_x}{m_y}&=\frac{2bB}{2aA}=\xi,
\end{align}where the lattice constants $a, b$ tend to zero, (maintaining $a \neq b$) and simultaneously the hopping elements $A, B$ tending to infinity. Since $A,B$ and the lattice constants $a,b$ are not equal in a rectangular lattice, the anisotropy parameter $\xi \neq 1$ causes the electron trajectory to be chaotic. Then, by applying a rapidly oscillating external field of the form $\textbf{f}=2(f_x,f_y) \cos{\omega t}$, the time--independent effective Hamiltonian function (\ref{8}) with $U(X,Y)=-U_1/(X^2+Y^2)^{1/2}$ is obtained, where $X,Y$ are the \textit{slow} position variables.

The results of the application of a homogeneous and rapidly oscillating external field in a rectangular lattice with the purpose of ``duplicating" it, have already been referred to in the footnote of p.\;2. Thus, it is known that with specific external fields one can modify the mass tensor components, such that the electron mass in the \textit{X} and \textit{Y} directions is the same, leading to stationary trajectories (quasi-ellipse in the lattice and ellipse in the continuum). In this regime chaotic motion disappears, suggesting a kind of chaos control. That is, the controlled manipulation of the components of the field $\textbf{f}=2(f_x,f_y) \cos(\omega t)$ which allows modifying the electron dynamics, so that stationary trajectories are possible when masses in the $X$,$Y$ directions are equal.

Consider next the effective masses on the $X,Y$ axes reparameterized by an external field of the usual type in this work, $\textbf{f} (\omega t)=2(fx,fy)\cos(\omega t)$. Thus, according to \eqref{10} and using in this case the effective Hamiltonian, we obtain the reparametrized components of the mass tensor:

\begin{align}
\label{13}
\tilde{m}_x&=\left[2Aa^2J_0(\alpha)\cos(aK_x(t))\right]^{-1},\nonumber\\
\tilde{m}_y&=\left[2Bb^2J_0(\beta)\cos(bK_y(t))\right]^{-1},
\end{align} where $\alpha = 2af_x/\hbar\omega$ and $\beta = 2bf_y/\hbar\omega$ are the arguments of the
Bessel functions. Here we see the dependence of the effective masses on the external electric field. The ratio of these components is

\begin{align}
\label{14}
\frac{\tilde{m}_x}{\tilde{m}_y}&=\xi \frac{\cos(bK_y)}{\cos(aK_x)}\frac{J_0(\beta)}{J_0(\alpha)},
\end{align}where $\xi \equiv b^2B/(a^2 A)$ is the mass anisotropy parameter with value $\xi=2.94$ in order to induce a chaotic trajectory around the impurity. The mass ratio in \eqref{14} will be equal to $\xi$ when $a=b=\alpha=\beta=0$, otherwise this ratio will depend on time. However, in the continuum limit given by $a, b\rightarrow 0$ and $A, B \rightarrow \infty$ such that $\xi$ is constant, it happens that $\cos(aK_x) \cong 1$ and $\cos(bK_y) \cong 1$, then the mass ratio can be approximated as

\begin{align}
\label{15}
\frac{\tilde{m}_x}{\tilde{m}_y} \cong \xi \frac{ J_0(\beta)}{J_0(\alpha)}.
\end{align}

For the purposes of this work, it is desired to maintain the reparameterized constant effective mass ratio equal to 1, which means that the masses become equal in the $X, Y$ directions. To achieve this the values of $\alpha, \beta$ must be chosen (numerically) such that

\begin{figure}[t]
\centering
\includegraphics[scale=0.4]{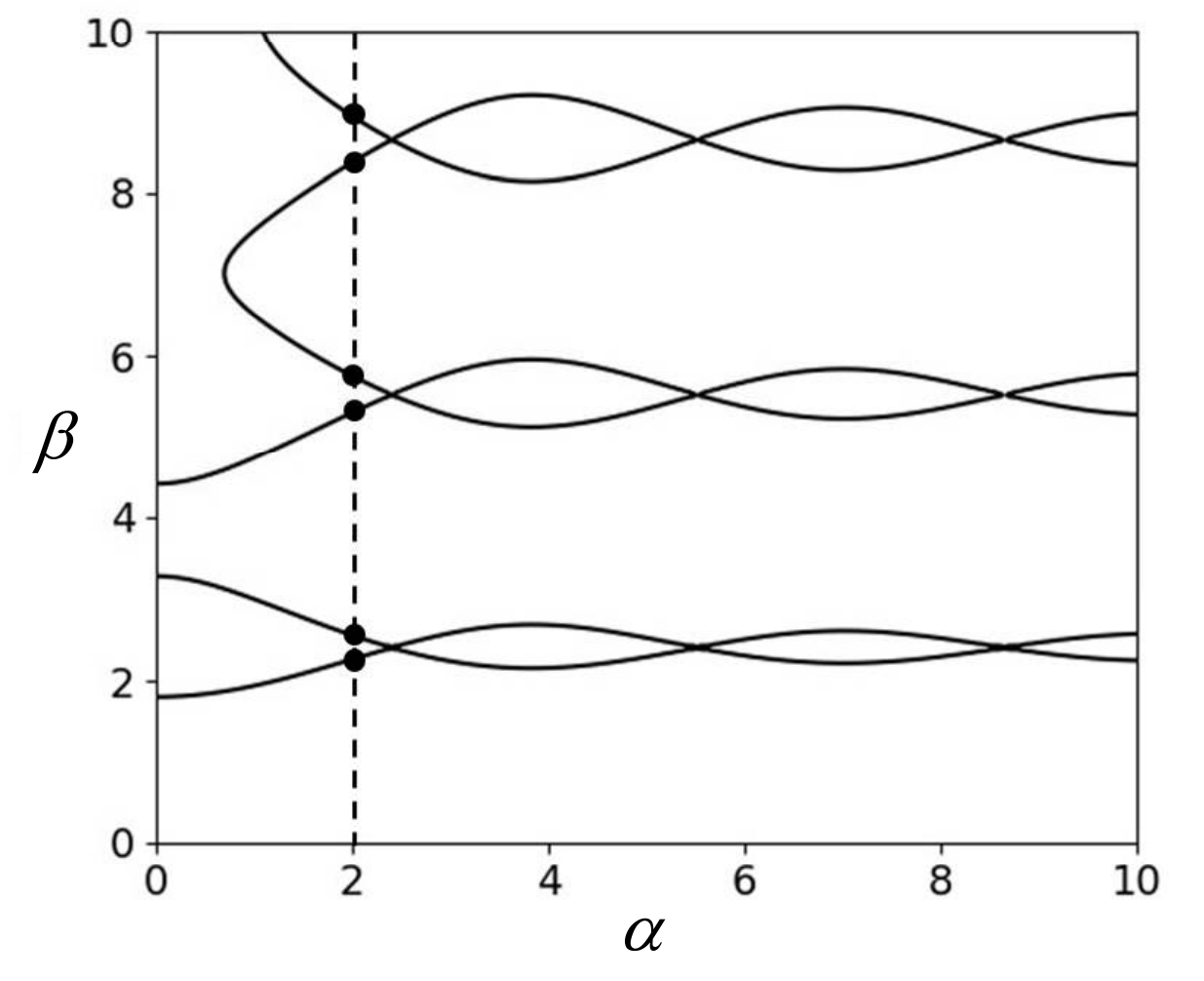}
\caption{Values of the external field parameters $\alpha, \beta$ defined in $\textbf{f}(\omega t)=(\alpha/a,\beta/b)\omega\cos(\omega t)$, with $a=20b$ and $\omega$ high enough such that the effective Hamiltonian \eqref{8} is valid (the case of Fig.\;\ref{fig11} below shows a numerical estimation of such a large value of $\omega$). The black dots correspond to the parameters for which the effective masses $m_x, m_y$ become equal as a consequence of the condition $J_0(\alpha)/J_0(\beta)=2.94$. The trajectories in this case are ellipses.}
\label{fig1}
\end{figure}

\begin{align}
\label{fig15a}
J_0(\alpha)/J_0(\beta)=\xi.
\end{align}This is represented in Fig.\;\ref{1} where different curves of possible combinations of $\alpha$ and $\beta$ that satisfy \eqref{fig15a} are shown. The values of $\alpha$ and $\beta$ contain the information of the amplitude and direction of the rapidly oscillating external field. In this way, the rapidly oscillating field will be:

\begin{align}
\label{16}
\textbf{f}(\omega t)=\left(\frac{\alpha}{a},\frac{\beta}{b}\right)\omega\cos{\omega t},
\end{align}with the lattice constants $a=20b$, maintaining this relation in the continuum limit and $\omega$ sufficiently high for the effective Hamiltonian to be valid. From here we note that the control parameters of the field \textbf{f} are given by the values of $\alpha$ and $\beta$, so that by choosing them, the dynamic behavior of the effective mass tensor of the electron is \textit{manipulated} (interactions engineering). The arbitrary choice of the rectangular cell dimensions $a=20b$ is due to the fact that it reproduces the results of Cort\'es in the continuum limit \cite{Cortes2008}; moreover, such a choice permits to obtain a desired result in the lattice, which is, the particle has less probability of hopping between nearest neighbours along the $X$ axis because $a \gg b$, yielding a vertical and oscillating trajectory as reported in \cite{Sanjines2025}.

\begin{figure}[t]
\centering
\includegraphics[scale=0.3]{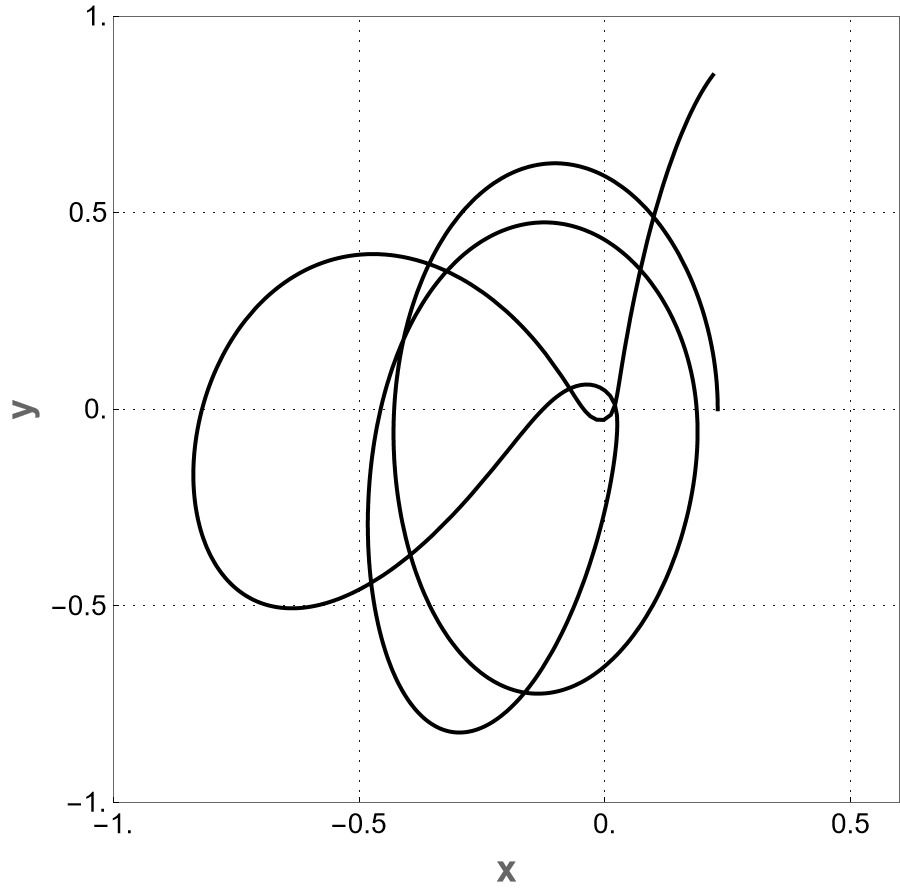}
\caption{Trajectory of the particle in the continuum with field parameters $\alpha=\beta=0$ and initial conditions $x(0)/a=230$ ($a$ is the length of the orthorrombic cell's largest side), $y(0)=0$, $k_x(0)=0$ and $k_y(0)=2.587$ (where the latter is in units of $1/a$ and was determined with \eqref{17}). This result reproduces that reported by Cort\'es \cite{Cortes2008}, wherein different masses $m_x,m_y$ on the $X,Y$ axes are introduced such that $\xi \equiv m_x/m_y=2.94$.}
\label{fig3}
\end{figure}

\begin{figure}[t]
\centering
\includegraphics[scale=0.5]{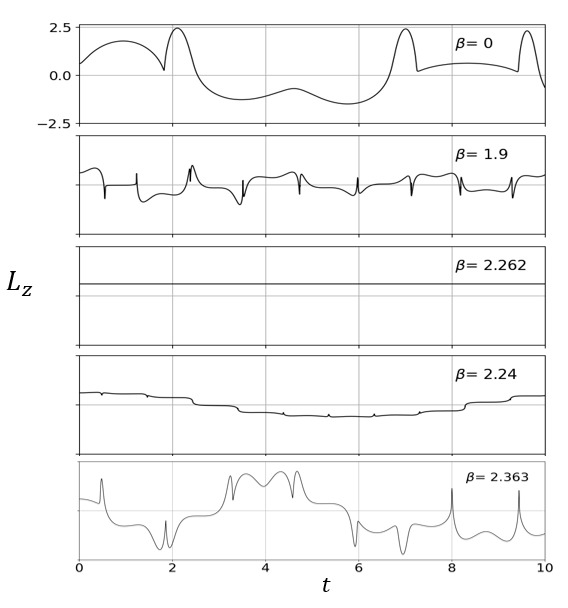}
\caption{Angular momentum corresponding to the trajectories with $\beta=\{0; 1.9; 2.24; 2.262;2.363\}$ from Fig.\;\eqref{fig5}. It can be seen that the angular momentum for these trajectories is irregular, except for $\beta=2.262$, where the trajectory is elliptical and its angular momentum is conserved}
\label{fig4}
\end{figure}

\begin{figure}[t]
\centering
\includegraphics[scale=0.45]{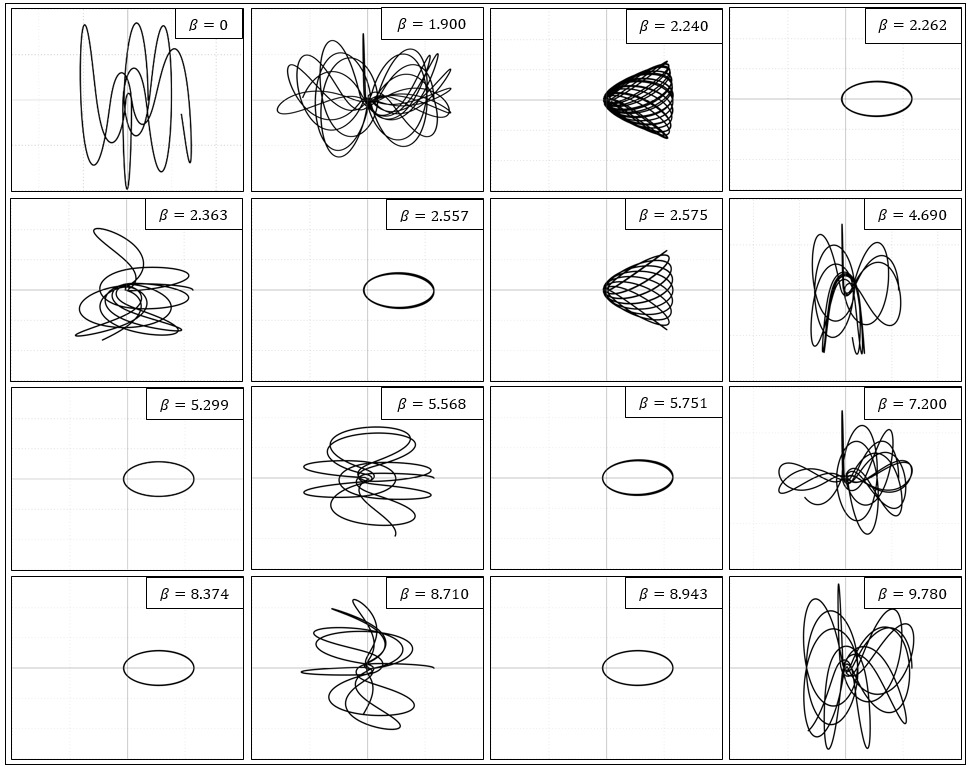}
\caption{Sequence of trajectories corresponding to the regions marked with circles in Fig.\;\ref{fig6}, where the chaos--order--chaos transition can be clearly seen; the ``order" or stationary regime includes the elliptical and quasi-elliptical trajectories (ellipse with precession), indistinguishable in region 3-4 of Fig.\;\ref{fig6}.}
\label{fig5}
\end{figure}

\section{Chaos control in the continuum}

We consider the motion of the particle (electron) initially at rest at a distance $r_0$ from the origin. From \eqref{1}, its total energy will be $H=E=-U_1/r_0$; that is, for $r=r_0$ there will only be potential energy and its kinetic energy is zero. In this way, the particle will orbit around the impurity in a circular region of radius $r_0$. Since the Hamiltonian in \eqref{1} does not explicitly depend on time, the total energy in this system is conserved \cite{Landau}; that is, for each point within the region $r \leq r_0$ the total energy will remain constant.

In this way we can assign an arbitrary value of $U_1$ in the potential energy $U=-U_1/r$ and an energy total $E$ so that the initial dimensionless position let $r_0=-U_1/E=1$, that is, the coordinates $x,y$ vary between $[-1,1]$. The particle is initially located at $x_0$ with a quasimomentum $k_y(0)$ along the $Y$ axis. The initial quasimomentum for the energy to remain constant is given by:

\begin{align}
\label{17}
k_y(0)=\frac{1}{b} cos^{-1}\left[ 1-\frac{1}{2B}\left( \frac{U_1}{x_0}-1 \right) \right],
\end{align} this expression can be applied in the continuum limit.

\begin{figure}[t]
\centering
\includegraphics[scale=0.6]{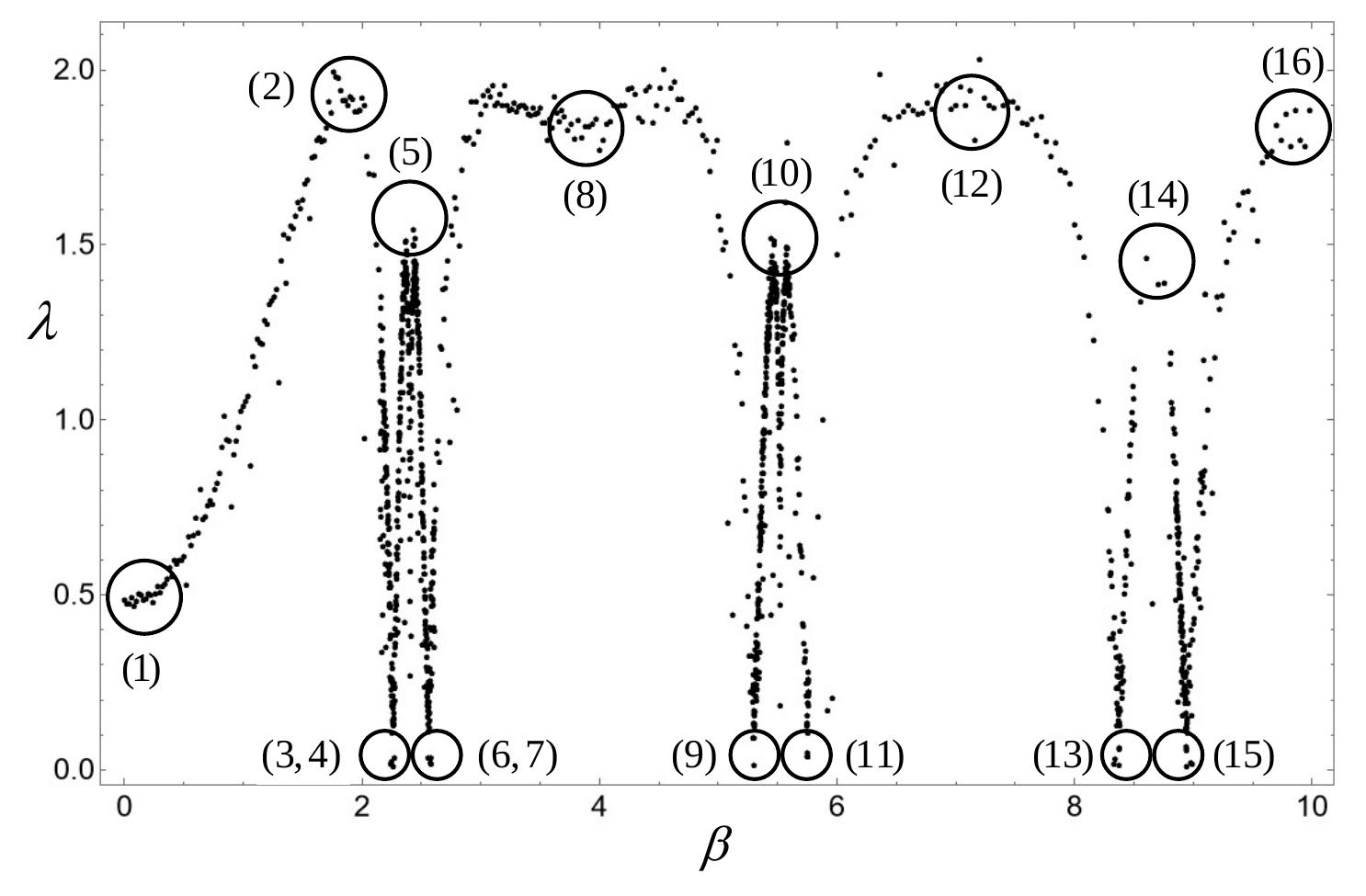}
\caption{Value of the Lyapunov exponent $\lambda$ vs. $\beta$, the vertical component  of the external electric field (with horizontal component $\alpha$ constant) for more than 1000 points. Specifically, the values of $\lambda$ and $\beta$ in the regions marked with the circles correspond, sequentially, to the trajectories of Fig.\;\ref{fig5}, where the transition between chaos and order can be observed. This is the phenomenon of \textit{chaos control}.}
\label{fig6}
\end{figure}

\begin{figure}[t]
\centering
\includegraphics[scale=0.4]{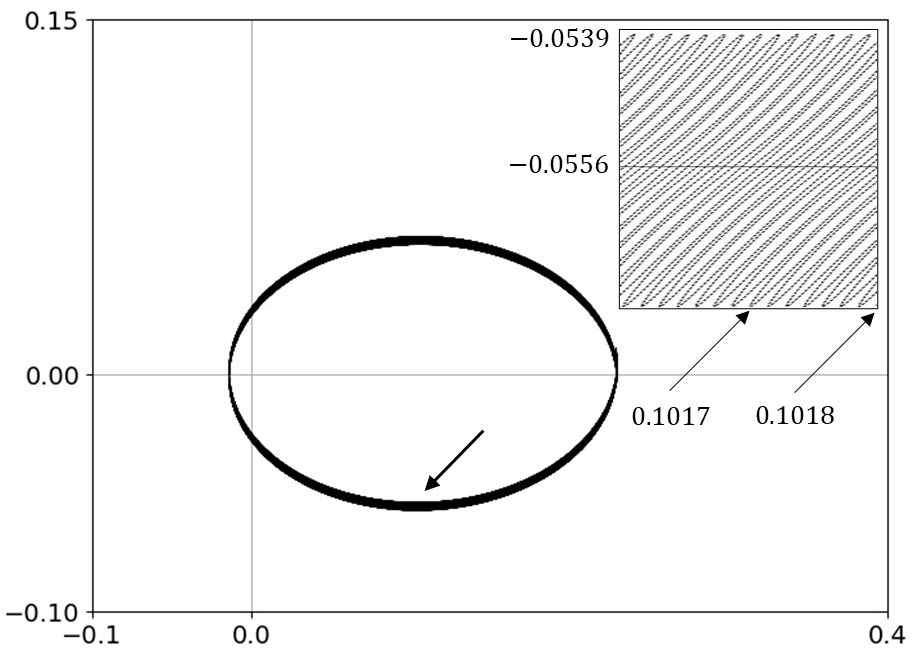}
\caption{Elliptical trajectory obtained from the transformed Hamiltonian $H(\textbf{r},\textbf{k}'(t))$ after (\ref{1a}) with a rapidly oscillating field $\textbf{f}(\omega t)=(\alpha/a,\beta/b)\omega\cos(\omega t)$ and values of $\alpha, \beta$ corresponding to chaos control in the continuum (Fig.\;\ref{fig1}). The condition $\omega \gg 2\pi/T$ such that the effective Hamiltonian $H_{ef}$ in (\ref{8}) is valid is numerically taken as $\omega/(2\pi/T)=2314$ in this simulation, where $T$ is the period of the trajectory calculated from $H_{ef}$. In the insert graphic the detailed rapid oscillations are shown for the small region indicated by the thick arrow; the horizontal solid line was calculated from $H_{ef}$ to verify that such line is -in fact- an element of the elliptic trajectory and represents very well the average of the rapid oscillations, as expected.}
\label{fig11}
\end{figure}

By choosing the control parameters of the field external electric $\alpha, \beta$ equal to zero, the external field is $\textbf{f}=(0,0)$. In this situation, the effective Hamiltonian \eqref{8} and the time-dependent Hamiltonian \eqref{1a} coincide with the strong coupling Hamiltonian \eqref{1}. This is because the Bessel function is 1 when its argument is zero, and also the amplitude of the external field is directly proportional to $\alpha$ and $\beta$. This implies that, under the initial conditions $x(0)=0.23$, $y(0)=0$, $k_x(0)=0$ y $k_y(0)=2.587$  (where this last quasimomentum was determined from \eqref{17}), the electron dynamics in the continuum should reproduce a chaotic trajectory consistent with the results already reported. In Fig.\;\ref{fig3} the corresponding trajectory is shown, which evidences that the system again reproduces the same chaotic behavior.

If values of the control parameters $\alpha$ and $\beta$ are chosen corresponding to the curves of Fig.\;\ref{fig1}, that is, those values of $\alpha , \beta$ that satisfy the condition $J_0(\alpha)/J_0(\beta)=2.94$, then the effective reparameterized masses will be equal in the $X,Y$ directions. Thus, it is expected that the particle trajectory is an ellipse, thereby achieving a kind of chaos control due to the rapidly oscillating external field. There are different ways to choose the values of $\alpha$ and $ \beta$. In this work a constant value $\alpha=2$ was chosen and different values of $\beta$ such that they represent the external electric field condition such that
$\tilde{m}_x/\tilde{m}_y=1$ (black dots in Fig.\;\ref{fig1}). In this range we note that we have 6 points representing this condition. Likewise, a constant value of $\beta$ could have been chosen and vary $\alpha$ instead in Fig.\;\ref{fig1}; we can see that there are also parameter values corresponding to chaos control. From the numerical evidence in Fig.\;\ref{fig1} when $\beta$ values are extrapolated beyond 10, there will not be intersections with $\alpha=2$, which suggests a finite number of pairs $(\alpha, \beta)$. This is the essence of engineering of interactions applied to Kepler's problem on the lattice in order to control chaos.

In Fig.\;\ref{fig4} the angular momentum is shown $L_z=XK_y-YK_x$ corresponding to the trajectories
of $\beta=\{0; 1.9; 2.24; 2.262;2.363\}$ of Fig.\;\ref{fig5} where it is clearly observed the transition from chaotic to the stationary regime (ellipses) for which the angular momentum is irregular or constant, respectively. In principle, for arbitrary values of $\beta$ the angular momentum for our system is not conserved, as the temporal variation of angular momentum $\dot{L}_z=(2a^2A-2b^2B)K_xK_y$ is caused by the anisotropy of the masses. On the other hand, for the value $\beta = 2.262$, angular momentum remains constant, as expected, because the effective masses are equal in both directions. Although the variation of angular momentum does not exhibit an apparent periodicity and the trajectory suggests chaotic behavior, the analysis of angular momentum alone is not sufficient to conclude that the system is chaotic. In Fig.\;\ref{fig5} the trajectories of the electron around the impurity for different values of the parameter $\beta$. It started with $\beta=0$, case in which the external electric field possesses only one component in the $X$ direction. As the value of $\beta$ increases, transitions between chaotic and elliptical trajectories are observed. The trajectories correspond to the value of $\alpha=2$ and the values $\beta=\{2.262; 2.557; 5.299; 5.751;8.374;8.943\}$ which correspond to the black dots of Fig.\;\ref{fig1}. For other values of $\beta$ not belonging to this set, the trajectories are chaotic, as observed in Fig.\;\ref{fig5}.

One way to characterize the sensitivity in the initial conditions of a chaotic system is through the Lyapunov
exponent $\lambda$ \cite{Strogatz}, which allows quantifying the exponential separation of trajectories as a function of time. Thus, trajectories with $\lambda>0$ are chaotic, trajectories with $\lambda<0$ are stationary (elliptical in this work) and the case of $\lambda=0$ corresponds to the quasi-stationary regime. In Fig.\;\ref{fig6}, the maximum Lyapunov exponent was calculated according to the code developed by Dubeibe F. L. in \textit{Mathematica} \cite{Dubeibe2013}. The value $\alpha=2$ was kept fixed while varying the value of $\beta$ (more than a thousand values). This resolution allows for a more detailed characterization of the system as a function of the control parameters.

We note that for the values $1.5<\lambda<2$ we have -in fact- chaotic trajectories. The values of $\beta$ for which the value of $\lambda$ decreases abruptly (close to zero) are those for which the effective masses are equal and the system presents regular elliptical trajectories. The regions marked with circles in Fig.\;\ref{fig6} correspond sequentially to the trajectories shown in Fig.\;\ref{fig5}, where the transition between chaos and order can be seen. Although is was not possible to obtain $\lambda \leq 0$ with this algorithm \cite{Dubeibe2013}, the elliptical trajectories Fig.\;\ref{fig6} can be theoretically ascribed to $\lambda<0$. Interestingly, the trajectories in Fig.\;\ref{fig5} which are neither chaotic nor elliptical have a very small value of $\lambda$ as shown in Fig.\;\ref{fig6}. These trajectories could be explained as follows.

Consider the variation of $J_0(\beta)/J_0(\alpha)$ as a function of $\beta$ (for $\alpha$ constant). When this ratio matches the value $1/\xi$, the ratio of the effective masses given by \eqref{15} is equal to 1. For $J_0(\beta)/J_0(\alpha) > 1/\xi $, the ratio of the effective masses is greater than 1; in consequence,$\tilde{m}_x$ will be greater than $\tilde{m}_y$. In the particular case $\beta=0$, the maximum value in \eqref{15} is obtained, so that the effective mass $m_x$ is significantly greater than $m_y$. Under these conditions, the particle shows a greater preference for moving in the $Y$ direction than in the $X$ direction. For this reason, we consider that the trajectory corresponding to this value of $\beta$ is ``less" chaotic. On the other hand, if $J_0(\beta)/J_0(\alpha) < 1/\xi $, the ratio of the effective masses in \eqref{15} is less than 1, that is, $\tilde{m}_x$ will be less than $\tilde{m}_y$, having a greater preference for moving along the $X$ axis than along the $Y$ axis, as can be seen in Fig.\;\ref{fig5} for the values $\beta = \{2.363;5.568;8.71\}$, which correspond to the regions marked with circles 5, 10, and 14 in Fig.\;\ref{fig6}.

With the results presented in this section, it can be seen that the interactions engineering scheme permitted us to modify the components of the mass tensor of the particle in the lattice (in the continuum limit), and consequently, to modify the particle's trajectories, showing the phenomenon of chaos control suggested in this work.

\section{Chaos control in the lattice}

\begin{figure}[t]
\centering
\includegraphics[scale=0.50]{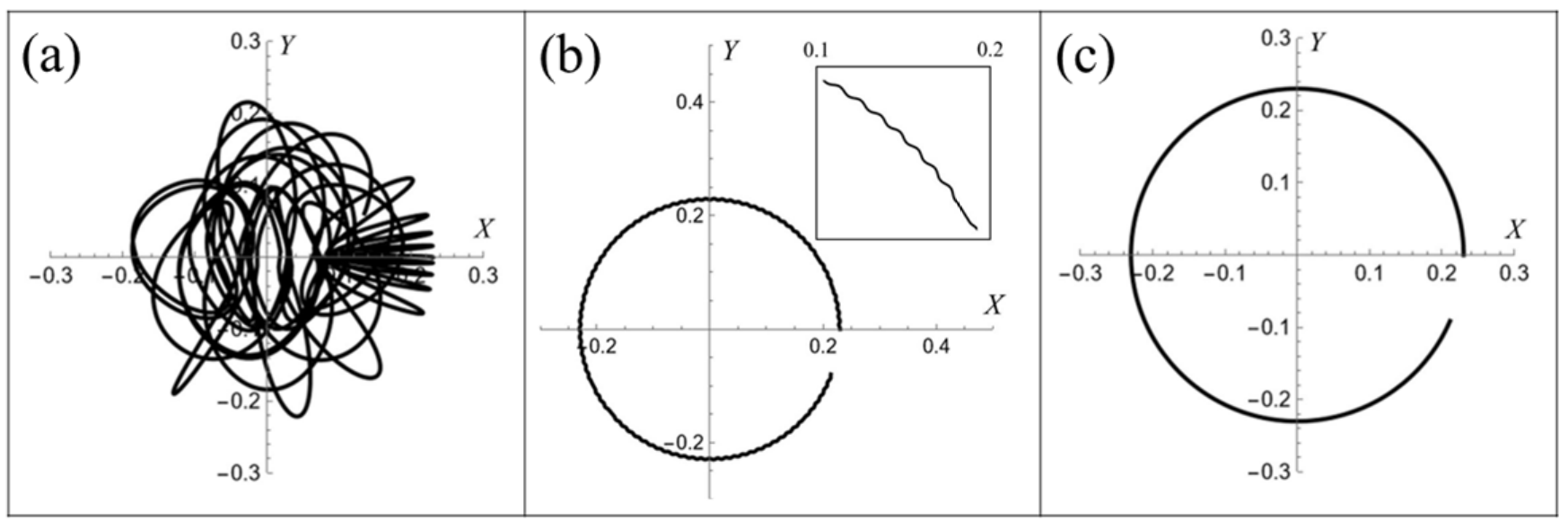}
\caption{Chaos control in the lattice: case of the circular trajectory. We use the parameters of Fig.\;\ref{fig3} with $\alpha, \beta$ corresponding to chaos control in the continuum (Fig.\;\ref{fig1}) and initial conditions $x(0)/a=1.15, y(0)=0$, $k_x(0)=0, k_y(0)=0.1$. With such values, the trajectory in (a) is chaotic. In (b) we solved (\ref{18})-(\ref{22}) for a circular trajectory $X^2+Y^2=R^2$. The result confirms such a circular trajectory; however, as we note in the insert graphic in (b), the element of arc has minute oscillations. The numerical value of the angular momentum (per unit mass) $l=X\dot{Y}-\dot{X}Y$ is approximately constant. Therefore, taking $l$ to be constant we deduce a constant and uniform magnetic field $B=U_1/(qRl)$. In (c) we introduced such a field in (\ref{18})-(\ref{21}) and solved them numerically. The result is a \textit{soft} circular trajectory without the oscillations observed in (b). In short, the difference between (b) and (c) is that in the latter $l$ is explicitly taken as a constant whereas in the former, it is not.}
\label{fig8}
\end{figure}

Let us now explore the mass ratio $\tilde{m}_x/\tilde{m}_y$ in (\ref{14}) in the rectangular lattice wherein the lattice constants $a, b$ are not negligible. In this case the term $\cos(bK_y)/\cos(aK_x)$ will be time-dependent. In order to have again, as in the continuum, that $\tilde{m}_x/\tilde{m}_y=1$, we propose the action of a weak external and uniform magnetic field $\textbf{B}(t)=\nabla \times \textbf{A} (t)$ perpendicular to the rectangular lattice.\footnote{We assume that such $\textbf{B}(t)$ field is weak enough so that it is not necessary to invoke the magnetic Bloch states (with a Berry curvature) as described, for example, by Chang M. and Niu Q. in \cite{Chang1995}} Incorporating the magnetic Lorentz force in the equations of motion for the effective coordinates and momenta $X, Y, K_x, K_y$, these are

\begin{align}
\dot{X}&=2\widetilde{A}a\sin{aK_{x}},\label{18}\\
\dot{Y}&=2\widetilde{B}b\sin{bK_{y}},\label{19}\\
\dot{K}_{x}&=-\frac{\partial U}{\partial X}+qB(t)\dot{Y},\label{20}\\
\dot{K}_{y}&=-\frac{\partial U}{\partial Y}-qB(t)\dot{X}.\label{21}
\end{align}These set of equations accounts for the Peierls phase $\exp (iq \int_i^j \textbf{A}\cdot d\textbf{r})$ which has to multiply the tight-binding elements $t_{ij}=\widetilde{A}, \widetilde{B}$ in the X, Y directions, respectively. In turn, the introduction of the Peierls phase is equivalent to defining a new effective Hamiltonian $H_{ef} (X,Y;\textbf{K}) \rightarrow H_{ef} (X,Y;\textbf{K}+q\textbf{A})$ where the kinetic momentum $\hbar \textbf{K}$ has been replaced by the canonical momentum $\hbar \textbf{K}_c \equiv \hbar \textbf{K}+q\textbf{A}$ (recall that $\hbar=1$ in this work and $q=-e$ for the electron). Then, the new semiclassical equations of motion are the Onsager equations: $\textbf{v}=\partial_{\textbf{K}_c} H_{ef} (\textbf{K}_c)$ for the velocity and $\partial_t\textbf{K}_c=-q \textbf{v}\times \textbf{B}$ for the Lorentz force. In this work we have added the Coulomb force to the set of equations (\ref{18})-(\ref{21}).

Now, we set about finding the specific value of $B(t)$ such that $\cos(bK_y)=\cos(aK_x)$ and therefore $\tilde{m}_x/\tilde{m}_y=1$ in (\ref{14}). That condition is equivalent to its time-derivative $b\dot{K}_{y}\sin(bK_y)=a\dot{K}_{x}\sin(aK_x)$, which combines with the equations (\ref{18})-(\ref{21}) to give

\begin{align}
B(t)=\frac{U_1}{q(a^2+b^2)}\left[\frac{a^2 X\dot{X}-b^2 Y\dot{Y}}{\dot{X}\dot{Y}(X^2+Y^2)^{3/2}}\right]\label{22}
\end{align}To test this expression we choose the very simple case of a circular trajectory given by $X^2+Y^2=R^2$ (or $X\dot{X}+Y\dot{Y}=0$) for which $(qR^3/U_1)B(t)=X/\dot{Y}=-Y/\dot{X}$ is obtained. Then, $B(t)\dot{Y}$ and $B(t)\dot{X}$ are readily substituted in (\ref{20}) and (\ref{21}), respectively. As expected, the numerical solution of (\ref{18})-(\ref{21}) gives a circular trajectory (Fig.\;\ref{fig8}). From this solution we can verify that the angular momentum (per unit mass) $l=|\textbf{r} \times \textbf{v}|=X\dot{Y}-Y\dot{X}$ is constant. Thus, $lB=XB\dot{Y}-YB\dot{X}=U_1/(qR)$ gives an static magnetic field $B(t)=B=U_1/(qRl)$ and a corresponding angular frequency $\omega=qB/(R^3 U_1)$ for the trajectory. One might be tempted to associate this result with a charged particle's classical circular trajectory with ciclotron angular frequency $\omega_c=qB/m$ in a uniform static magnetic field $B$. It is evident that $\omega \neq \omega_c$ since the physical approaches are different: in the lattice the electron moves according to semiclassical equations of motion with the quantum tight-binding band considered in (\ref{1}).

The next logical step is to solve (\ref{18})-(\ref{22}) for an elliptic trajectory $(X/R_x)^2+(Y/R_y)^2=1$ and take the angular momentum (per unit mass) $l$ as a constant. It is expected that the uniform magnetic field $B(t)$ should be time-dependent. We leave this case as an interesting issue to be studied elsewhere.

\begin{figure}[t]
\centering
\includegraphics[scale=0.60]{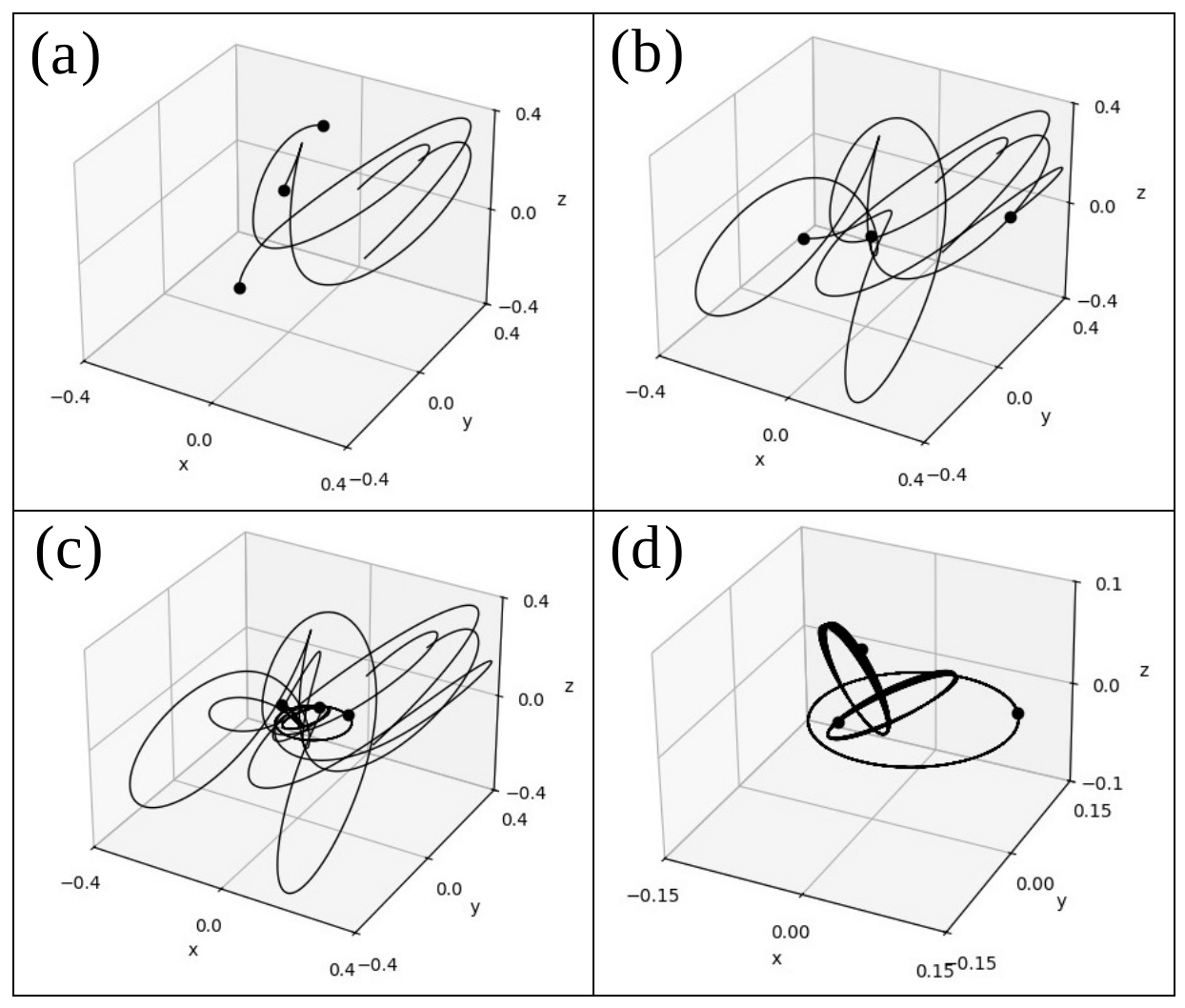}
\caption{Trajectories of 3 particles in a 3D orthorrombic lattice for the parameters: (a) $0<t<2, \beta=0.905$; (b) $0<t<4, \beta=1.809$; (c) $0<t<8, \beta=2.262$; (d) $8<t<15.5, \beta=2.262$ (in all cases $\alpha=2$). The particles' instantaneous positions are marked by the black points; the initial positions and velocities are different among them but are repeated in cases (a)--(d). An external and rapidly 3D oscillating electric field is applied according to (\ref{16}) extended in the $Z$ axis as $\textbf{f}(\omega t)=\left(\alpha/a,\beta/b,\beta/c\right)\omega\cos{\omega t}$. Interestingly, this $Z$-component has the same value of the $\beta$ parameter corresponding to the chaos control along the $Y$ axis. In cases (a) and (b), the values of $\beta$ do not yield chaos control whereas in (c) and (d), $\beta$ correspond to elliptic orbits. In (c) we observe that such orbits start emerging at the end of the time interval; for further clarity, in (d) we included a time-interval corresponding to the ellipses only.}
\label{fig9}
\end{figure}

\section{Chaos control in 3D: induced magnetic field}

In this section we show that two dynamical regimes could be related to the induced magnetic field due to a classical electron gas upon the site of the impurity ion as a consequence of Biot-Savart law. In this model the tight-binding electrons move in the orthorrombic lattice by the hopping mechanism subject to the superposition of the impurity Coulomb field and the rapidly oscillating and uniform external field, this latter being responsible of the chaos control in the continuum. Consistently to Bohr-van Leeuwen theorem \cite{Huang}, both regimes yield a net induced magnetic field $\langle\textbf{B}\rangle=0$ due to the classical electron gas in thermal equilibrium; however, since those regimes are clearly different, it might be possible -in principle- to distinguish one from the other by an observable property.

\begin{figure}[t]
\centering
\includegraphics[scale=0.4]{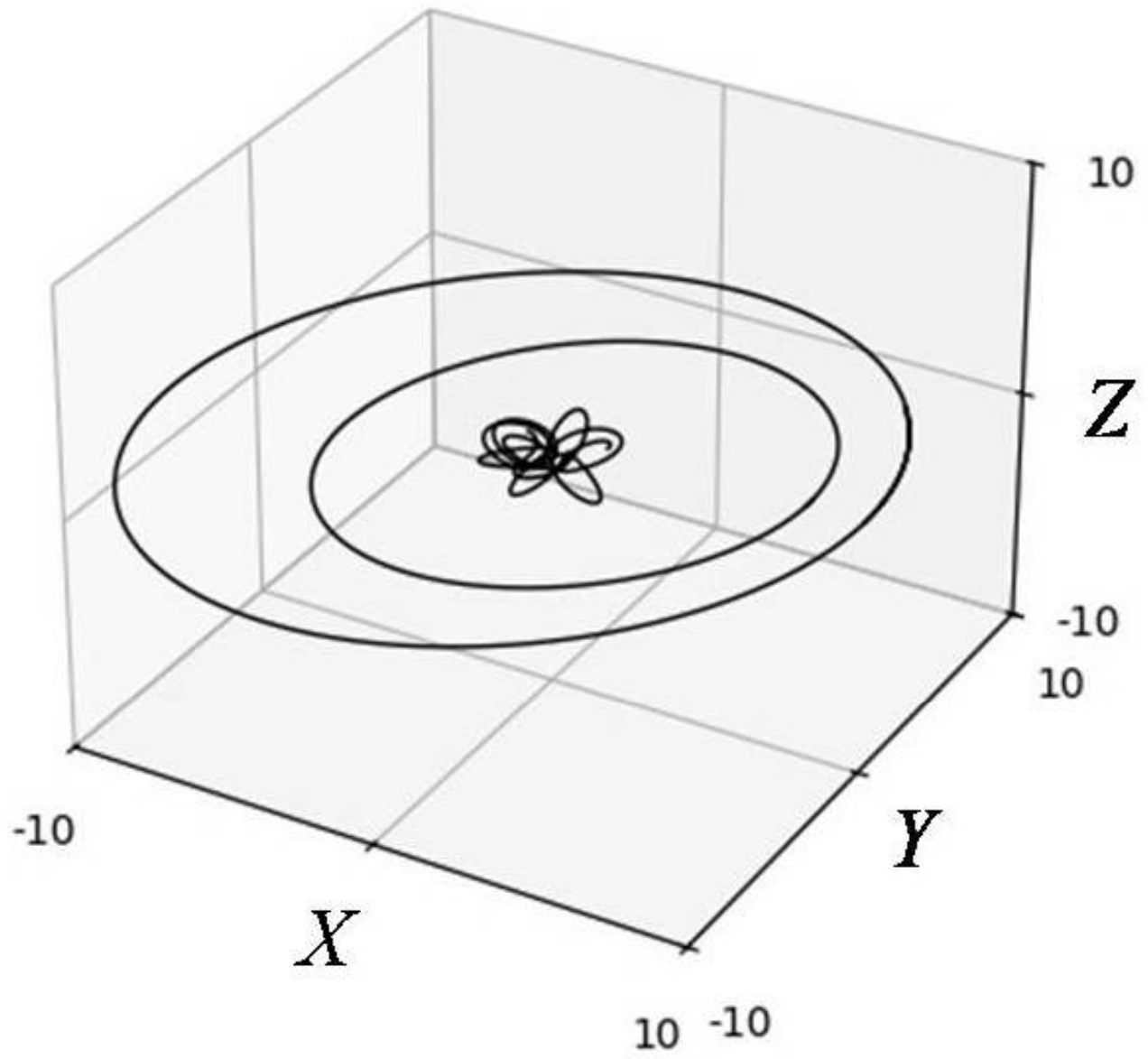}
\caption{Chaotic and stationary (elliptic) trajectories for values of $\alpha=2$, $\beta=2.262$, a quasimomentum with magnitude $k_0=2.587$ and the initial positions along the X axis: $x/a=2$, $x/a=9$ and $x/a=11$. The axes scales are expressed in units of $a$, the largest lattice constant of the orthorrombic lattice. We can thus distinguish between a ``near" region with a chaotic trajectory and a ``far" region with stationary trajectories.}
\label{fig12}
\end{figure}

We start by calculating the variable magnetic field $\textbf{B}(t)$ produced by a charged particle with charge $q$ located instantly at $\textbf{r}=(X,Y,Z)$ upon the site of the impurity located at $\textbf{r}=0$ (the trajectories of 3 particles are shown in Fig.\;\ref{fig9}). We use the expression $\textbf{B}(t)=(\mu_0 q/4 \pi)\;\textbf{r}\times \dot{\textbf{r}}/r^3$ (referred to in Jackson \cite{Jackson}) which gives the components
\begin{align}
B_i(t)=\frac{\mu_0 q}{4 \pi} \; \frac{X_j\dot{X_k}-X_k\dot{X_j}}{(X^2+Y^2+Z^2)^{3/2}} \label{23}
\end{align}with $(i,j,k)$ a cyclic permutation of $(1,2,3)$ and $(X_1,X_2,X_3)=(X,Y,Z)$. As we can see in Fig.\;\ref{fig9}, there will occur two posible dynamical regimes: the chaotic one (Fig.\;\ref{fig9}(b)) and the stationary one (Fig.\;\ref{fig9}(d)). In Fig.\;\ref{fig12} we observe the same regimes but obtained in this case by varying the initial positions instead of varying the $\alpha, \beta$ parameters, which stay fixed with the values corresponding to the chaos control in the continuum. This means that a particle whose initial position is ``far" from the origin moves as it does in the continuum limit of the lattice, that is, the lattice looks a quasi-continuum space. In the other hand, if the particle`s initial position is close to the origin, the space looks discrete and the $\alpha, \beta$ parameters do not yield chaos control, thus the trajectories are chaotic. Now, we propose a physical scenario: a classical gas of charged spinless particles move randomly in thermal equilibrium around the impurity site at the origin; by the action of the external field $\textbf{f}(\omega t)=\left(\alpha/a,\beta/b,\beta/c\right)\omega\cos{\omega t}$, the particles initially located near the impurity will fall into a chaotic regime, while the particles located ``far" from the impurity will fall into a stationary regime due to the chaos control technique.

We have made simulations for a model of an ideal gas of 1000 particles initially distributed randomly on a spherical shell of radius $R_0$ and whose center is the impurity site. Their velocities also have directions distributed  randomly but with a constant magnitude $v_0=\sqrt{3 k_B T/m}$, consistent with an ideal gas which is in thermal equilibrium with its surroundings at an absolute temperature $T\sim 300$ K. This gas could be thought of as the fraction of electrons in the conduction band of a doped semiconductor that were thermicaly promoted from the valence band across a gap of few meV. Thus, the gas is bounded around the donor impurity (a positive ion at rest) by the Coulomb interaction; the electrons in the gas move by the hopping mechanism in the semiconductor lattice. A typical distance between donor impurities in a doped semiconductor was taken as 100 nm, thus, the radius of the spherical surface is $R_0\sim 50$ nm.

\begin{figure}[t]
\centering
\includegraphics[scale=0.5]{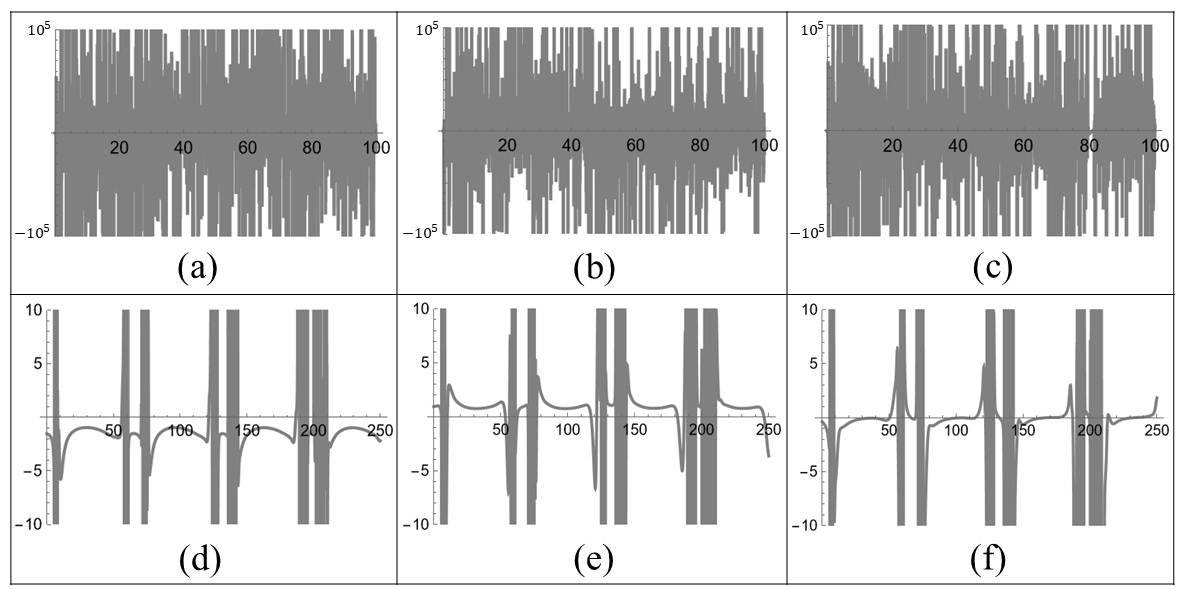}
\caption{Components $B_x, B_y, B_z$ of the induced magnetic field $\textbf{B}(t)$ at the impurity site due to 1000 charged particles (electrons) moving around it and initially located randomly on a sphere of adimensional radius $R_0/a$ expressed in multiples of the maximum lattice constant $a$ of the orthorrombic lattice. Cases (a), (b) and (c) correspond to $R_0/a=1$, which yield chaotic trajectories analogous to those of Fig.\;\ref{fig9}(b). Cases (d), (e) and (f) correspond to $R_0/a=10$, which yield elliptic trajectories analogous to those of Fig.\;\ref{fig9}(d). The very rapid and abrupt oscillations in each regime correspond to very small distances between the impurity and the particle`s instant positions according to (\ref{23}). For the elliptic trajectories those abrupt oscillations correspond only to those particles which pass by near the impurity due to their initial velocities oriented along the radial direction; the ``doublet" structure of these oscillations occurs because of the opposite directions of those velocities. The average values of $\textbf{B}(t)$ for the time-scale shown in the horizontal axes are $\langle\textbf{B}\rangle=(0.010,0.004,-0.004)\times 10^5$ and $\langle\textbf{B}\rangle=(-1.27,0.87,0.75)$ for cases (a), (b), (c) and (d), (e), (f), respectively. These values can be considered as negligible as compared to the values of the vertical axes scales. We can therefore take as granted the limit $\langle\textbf{B}\rangle=0$ for a very large number of particles.}
\label{fig10}
\end{figure}

In Fig.\;\ref{fig10} we show the components $B_x, B_y, B_z$ of the induced magnetic field $\textbf{B}(t)$ at the impurity site due to the gas particles moving around it; the shell's radius was expressed in adimensional units of $R_0/a$ where $a$ is the maximum lattice constant of the orthorrombic lattice. It can be observed a substantial physical difference between the chaotic and the stationary regimes: for the former there are not closed and stationary trajectories, the total $\textbf{B}$ components oscillate randomly and consequently there is not a net induced magnetic field at the impurity site; for the latter, a single elliptic trajectory induces a magnetic field at the impurity site but the the total components of $\textbf{B}$ due to all the randomly oriented moments within the shell are clearly distinguished in Fig.\;\ref{fig10} by almost plain regions separated from rapidly oscillating regions, giving thus an average $\langle\textbf{B}\rangle=0$, consistently with Bohr-van Leeuwen theorem.

The magnitude of the induced magnetic moment of a particle (charge $q=-e<0$) in a elliptical trajectory around the impurity (charge $Q>0$) is $M=iA=(q/T)A$, where $T=\pi K|Qq|\sqrt{m/(2|E|^3)}$ is the period of the trajectory and $A=\pi K|QqL|/\sqrt{8m|E|^3}$ is its enclosed area. Then, $M=q|L|/2m=q|R_0 \times v_0|/2=(q R_0 v_0/2)\sin \phi$ with $0<\phi<\pi$. The average value $\langle M \rangle$ for all possible directions $\phi$ is thus compared with the Bohr magneton as $\langle M \rangle/\mu_B=4mR_0v_0/(\pi \hbar)$. A rough estimation is this magnitude would be $\langle M \rangle/\mu_B\sim 10^2$, which tells us that the average effect of a single stationary (elliptic) trajectory can not be neglected as compared with typical atomic magnetic effects.

\section{Conclusions}

In this work, we studied, under the semiclassical approximation, the dynamics of a charged particle (electron) moving by the hopping mechanism in a orthorrombic lattice in the presence of a static Coulomb field (impurity) and a rapidly oscillating homogeneous external field. For its projection in a rectangular lattice in the continuum limit, it was possible to give a physical explanation of the mass anisotropy parameter $\xi=m_x/m_y= 2.94$ associated to chaotic trajectories.

We have shown that the application of an external and rapidly oscillating electric field is the mechanism to manipulate the effective masses and thus modify the dynamics of the particle. To achieve this, we found that the electric field must be of the form $\textbf{f}(\omega t)=(\alpha/a,\beta/b)\omega\cos(\omega t)$, where $\alpha$ and $\beta$ are the control parameters, which were appropriately chosen so as to modify the effective masses $\tilde{m}_x$ and $\tilde{m}_y$ to become equal. In this case the trajectories are ellipses in the continuum limit of the rectangular lattice, while for other values of $\alpha$ and $\beta$ corresponding to $\tilde{m}_x/\tilde{m}_y \neq 1$ the trajectories are chaotic. By fixing the value of $\alpha$ and varying the value of $\beta$ continuously, we have shown transitions from chaotic to stationary regimes (ellipses) by means of their Lyapunov exponents, thus consistently confirming the chaos control by means of the interaction engineering technique.

An heuristic extension of the chaos control technique to a rectangular lattice without invoking the continuum limit was proposed by the action of an external homogeneous magnetic field perpendicular to the lattice. It was found that a chaotic trajectory could be turned into a circular one by solving numerically the semiclassical Onsager equations valid for a tight-binding charged particle moving in a weak magnetic field.

An extension of the chaos control in the 3D orthorrombic lattice was made for a classical gas of 1000 spinless charged particles in thermal equilibrium and moving around the impurity site, giving the expected result $\langle\textbf{B}\rangle=0$ (Bohr-van Leeuwen theorem). However, it was also possible to observe two different dynamical regimes, the chaotic and the stationary ones, which might suggest that an observable property would be worth looked for to distinguish one from the other.

As a possible perspective of this work we suggest investigating what would be the chaos control conditions for a doped semiconductor with a lattice structure other than orthorrombic, for example, for the case of the silicon doped with arsenic (diamond structure) and see if it is possible to apply the chaos control scheme. Other perspective is to investigate if the dynamical regimes calculated numerically in Fig.\;\ref{fig10}, specifically cases (d), (e), (f), could be distinguished -in principle- by the bremsstrahlung radiation emitted from electrons passing by too near the impurity or other type of dispersive phenomena.

\end{document}